\documentclass[aps,prl,twocolumn,letterpaper,superscriptaddress]{revtex4-1}
\usepackage[caption=false]{subfig}
\pdfoutput=1
\usepackage{times}
\usepackage[pdftex]{graphicx}

\begin{document} 

\title{Solving Structure with Sparse, Randomly-Oriented X-ray Data}

\author{Hugh T. Philipp}
\affiliation{Cornell University, Laboratory of Atomic and Solid State Physics, Ithaca, NY USA.}
\author{Kartik Ayyer}
\affiliation{Cornell University, Laboratory of Atomic and Solid State Physics, Ithaca, NY USA.}
\author{Mark W. Tate}
\affiliation{Cornell University, Laboratory of Atomic and Solid State Physics, Ithaca, NY USA.}
\author{Veit Elser}
\affiliation{Cornell University, Laboratory of Atomic and Solid State Physics, Ithaca, NY USA.}
\author{Sol M. Gruner}
\affiliation{Cornell University, Laboratory of Atomic and Solid State Physics, Ithaca, NY USA.}
\affiliation{Cornell's High Energy Synchrotron Source (CHESS), Ithaca, NY  USA.}

\begin{abstract}
Single-particle imaging experiments of biomolecules at x-ray free-electron lasers (XFELs) require processing of hundreds of thousands (or more) of images that contain very few x-rays. Each low-flux image of the diffraction pattern is produced by a single, randomly oriented particle, such as a protein. We demonstrate the feasibility of collecting data at these extremes, averaging only 2.5 photons per frame, where it seems doubtful there could be information about the state of rotation, let alone the image contrast. This is accomplished with an expectation maximization algorithm that processes the low-flux data in aggregate, and without any prior knowledge of the object or its orientation. The versatility of the method promises, more generally, to redefine what measurement scenarios can provide useful signal in the high-noise regime.
\end{abstract}

\maketitle 

The ultra-intense, ultra-fast x-ray pulses from x-ray free electron lasers (e.g., the Linac Coherent Light Source, LCLS), hold potential to provide structural information about proteins that are not available in crystalline form~\cite{neutze2000}. Even with XFELs the number of photons scattered at large angles from a single protein is, on average, much less than one photon per pixel per image (i.e. frame). Since the measurement is destructive, many images must be gathered, each from a new molecule.  The situation is further complicated because the samples are imaged in unknown, random orientations.  The combination of low-flux and unknown particle orientation tests the information limits posed by measurements with few photons~\cite{elser2009}.

 Recovery of detailed, orientation specific structural information from a data set using many ultra-low intensity x-ray images from a randomly oriented sample has yet to be demonstrated. The information contained in these snapshots with only a few photons per frame would at first seem deficient to contain both a description of the orientation and intensity of the underlying structure. We demonstrate a simplified version of an algorithm developed for the x-ray single particle imaging experiment~\cite{loh026705,Loh2010} is capable of reducing this kind of data even with images that average only 2.5 x-ray photons per frame ($\sim 10^{-4}$ photons per pixel per frame). For this demonstration, we use a pixel array detector having the same CMOS front-end as the instrument currently installed at the Coherent X-ray Imaging (CXI) beamline at the Linac Coherent Light Source (LCLS). ~\cite{Philipp2010,PhilippJinst1}

The algorithm for reconstructing the x-ray intensity from data follows the expectation maximization (EM) principle~\cite{baum1970}.  EM starts with a random model of the intensity,  $w(i)$, at each pixel $i$.   These values are iteratively updated by a rule that can only increase the likelihood of the model.  The initial model is random because no information about the model is known. 

Each iteration involves two steps. In the first step, each frame of data, $f$, is assigned a probability distribution, $p_f(r)$, with respect to its unknown rotation, $r$, relative to the current intensity model. The rotations are sampled in increments of $2\pi/N$, where $N$ defines the angular resolution of the reconstruction. Each frame comprises photon occupancy, $k_f(i)$, at pixel $i$, which in our low-flux experiment are almost all zero, the exceptions being equal to 1. Because the photon counts are independent Poisson samples of the intensity at each pixel, the probability is 

\begin{equation}
p_f(r)\propto \prod_i \frac{w(i+r)^{k_f(i)}}{k_f(i)!}e^{-w(i+r)}\propto \prod_{i\in I_f} w(i+r),
\label{eqn:step1}
\end{equation}
where $i+r$ is rotation $r$, applied to pixel $i$, $I_f$ is the set of pixels recording photons in frame $f$, with the final expression applying in the low-flux limit. After normalizing the distributions, $p_f(r)$, the algorithm proceeds to the second step. 

In the second step the algorithm aggregates the photon data from all the frames according to the distributions obtained in the first step: 
\begin{equation}
w'(i)=\left\langle\sum_{r} p_f(r) k_f(i-r)\right\rangle_f.
\label{eqn:step2}
\end{equation}
The updated intensity model $w'(i)$ is an average of the photon counts in all frames with the appropriate distribution of rotations applied to each one. Interpolation is used in both steps, when mapping one square grid (model) onto one that has been rotated (detector). The EM algorithm is still valid when the data is winnowed by a structure-neutral criterion, such as the photon occupancy. In our implementation, for example, we discarded all frames with zero occupancy. 

This algorithm is closely analogous to the EMC algorithm that was developed for reconstructing the 3-dimensional (3D) intensity of randomly oriented particles~\cite{loh026705}. A 3D reconstruction is technically challenging because (i) the space of rotations is three-dimensional and (ii) the data is tomographic in nature, each frame providing information within only one 2D slice of the 3D model. And while the work required for the 3D reconstructions is correspondingly greater, we find that the performance of the algorithm in 3D with simulated data, and the 2D reconstructions reported here with actual data, is virtually the same when evaluated in terms of convergence (iteration number) in the ultra-low-flux limit. Thus,  our experiment is directly relevant to the case of single-protein imaging at an XFEL.

\begin{figure*}[!t]
\subfloat[]{\label{fig:realpadman1}
\reflectbox{\includegraphics[width=0.5 \columnwidth]{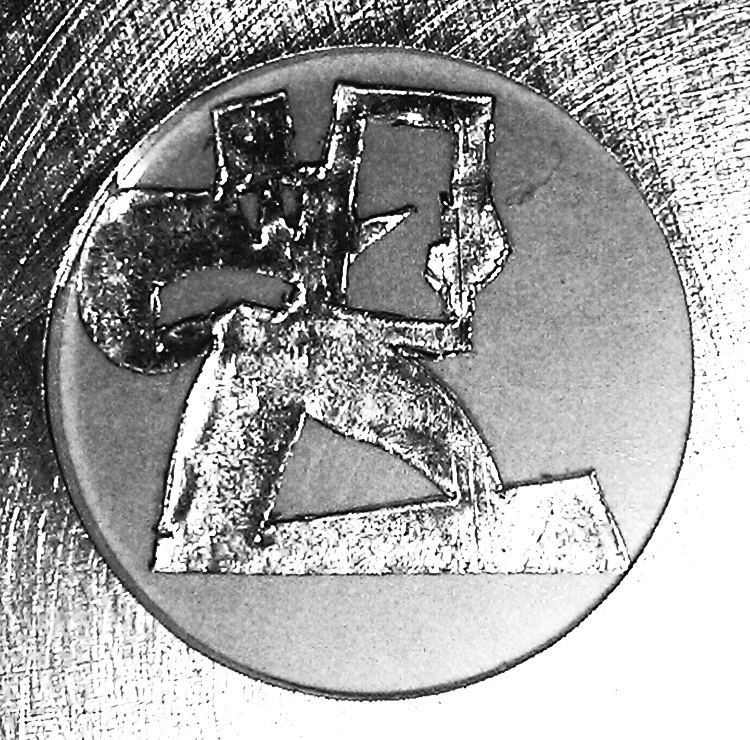}
} }
\subfloat[]{\label{fig:realpadman2}
\includegraphics[width=0.52 \columnwidth]{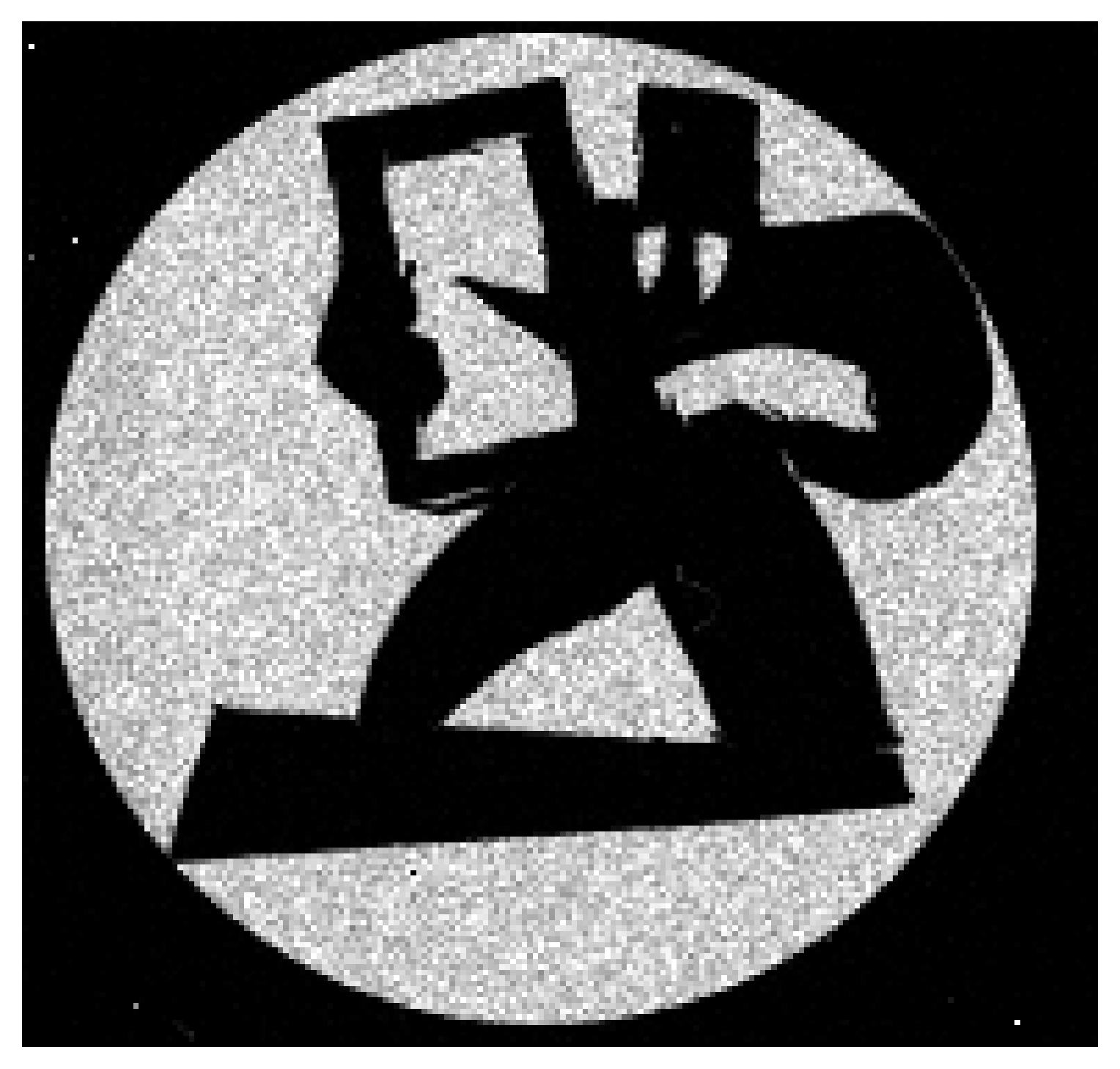}
}
\subfloat[]{\label{fig:115man}
\includegraphics[width=0.5 \columnwidth]{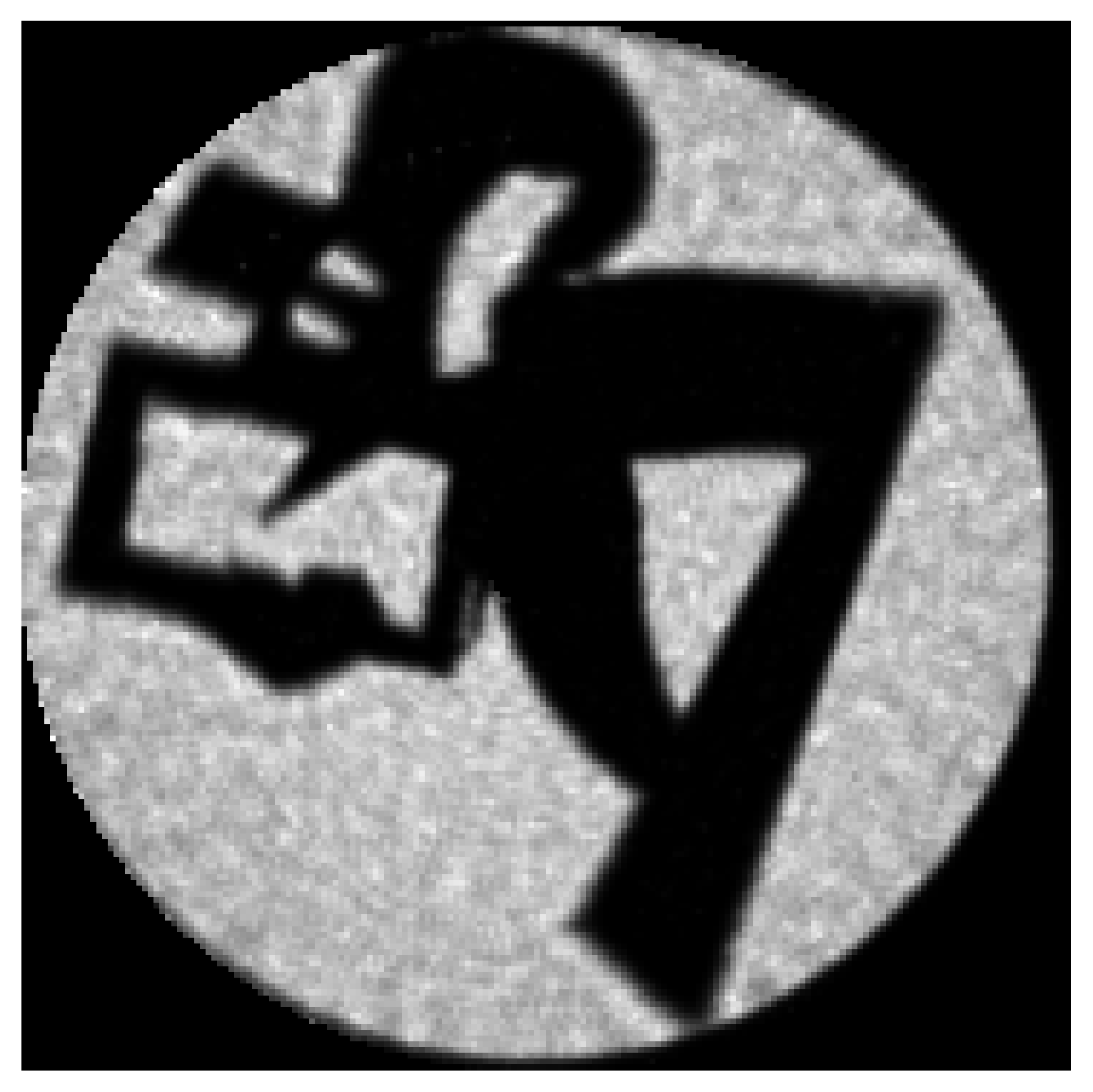}
}
\subfloat[]{\label{fig:25man}
\includegraphics[width=0.5 \columnwidth]{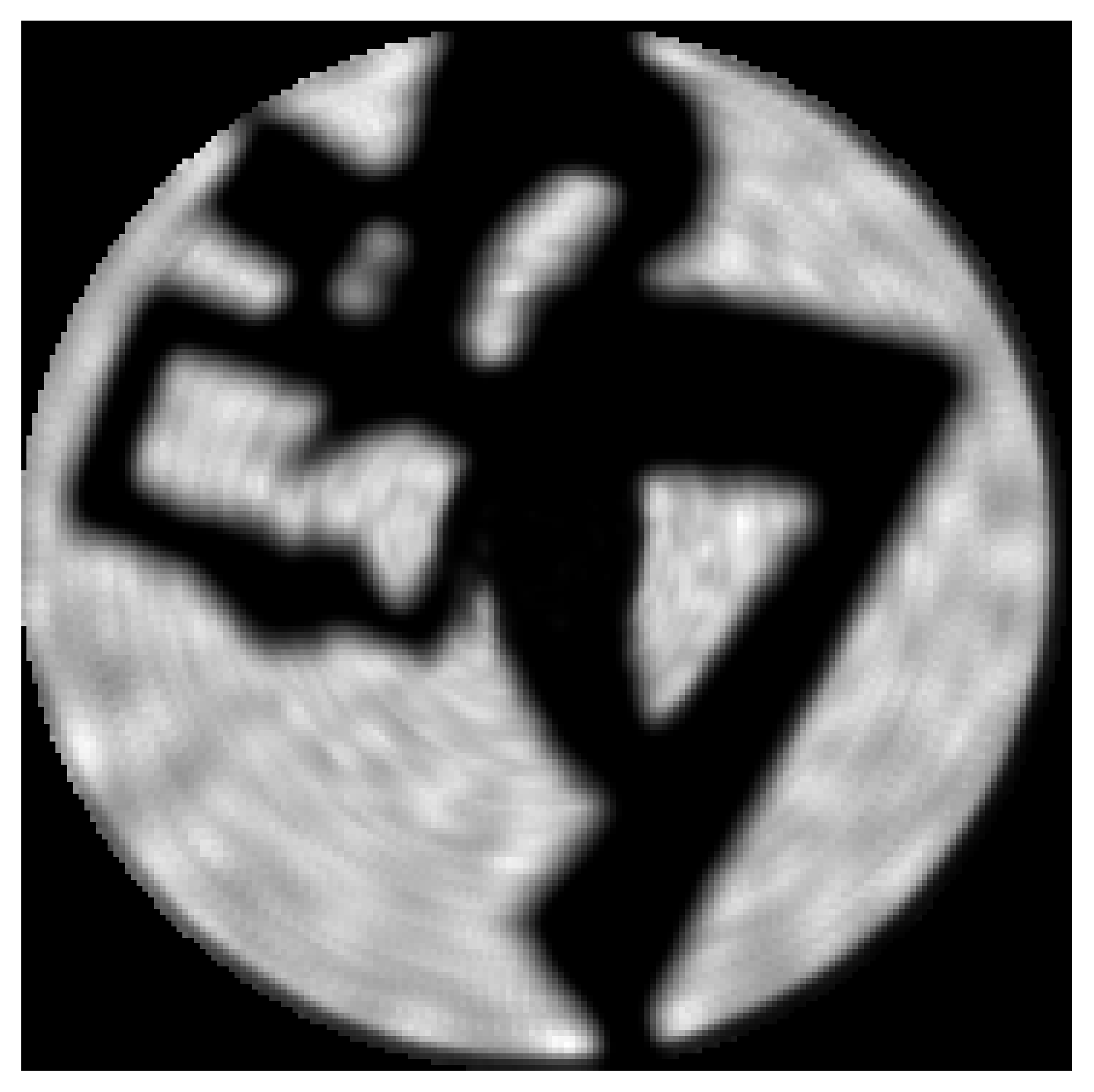}
}
\caption{(a) The lead x-ray mask mounted within an aperture in an aluminum disk.  
For (b), no rotations, approximately $3000$ photons/frame and 432 frames. For (c), random rotations, 11.5 photons/frame and $10^5$ frames and a total of approximately $1.2$ million recorded photons. For (d), random rotations, 2.5 photons/frame and $5 \times 10^5$ frames and a total of approximately $1.2$ million recorded photons.}
\label{fig:realpadman}
\end{figure*}

An alternative approach being considered~\cite{Huldt2003} for determining the rotations of randomly oriented particles involves classifying data on the basis of cross-correlations: 

\begin{equation}
c_{ff'}=\sum_i k_f(i) k_{f'}(i).
\end{equation}
This is not an option in the ultra-low flux limit because the numbers $c_{ff'}$ are essentially all zero, and in any event do not distinguish frames derived from like or unlike particle orientations. The EM algorithm, by contrast, compares each frame not with other frames but with the model. Greater sensitivity of rotation determination in the EM algorithm can be traced to the multiplicative nature of the comparison expressed by equation (\ref{eqn:step1}). 

The EM algorithm should in principle work with arbitrarily low-flux data.  It is clear that this is the case when we consider that there will be rare fluctuations where the photon occupancy is 2 or greater, even when the mean is just a fraction of a photon. A fair assessment of the viability of reconstructions in the low-flux regime must therefore take into account the inevitability of background.  The effects of background in the interpretation of our results are well captured by a simple information rate ratio: 

\begin{equation}
\displaystyle
R=
\frac{\sigma(1 + \mathrm{SN}^{-1}) \log{(1 + \mathrm{SN})} - (\sigma + \mathrm{SN}^{-1}) \log{(1 + \sigma\, \mathrm{SN})}}{-\sigma\log{\sigma}}.
\end{equation}
\noindent This is the information rate at signal-to-noise $\mathrm{SN}$ (the ratio of signal to background photons) divided by the rate in the limit of infinite $\mathrm{SN}$. The formula applies to signals that are two-valued, as in our experiment, where $\sigma$ is the open fraction of the mask. It is derived from the mutual information, $I(w,k)$, associated with the measurement of $k$ photons and an integrated flux, $w$, at one pixel. The joint probability distribution for this process is
\begin{equation}
p(w,k)=p(w)p(k|w),
\end{equation}
where $p(k|w)$ is the Poisson distribution for the photon count $k$ given flux $w$, and
\begin{equation}
p(w)=(1-\sigma)\delta(w-\nu)+\sigma\delta(w-\nu-\mu)
\end{equation}
describes the distribution of flux in our model. We take the limit where both $\mu$, the signal flux, and $\nu$, the background flux, are much less than one photon. The quantity $R$ is the mutual information $I(w,k)$ for the experimental parameters divided by the mutual information of the zero-background comparison, where $\nu$ is set to zero; the result depends only on the ratio $\mathrm{SN}=\mu/\nu$.

As an example, consider the case of $\mathrm{SN}=1/10$, which for $\sigma=0.6$ gives $R\approx 0.01$. A low flux experiment with 2 signal photons per frame and this poor signal-to-noise would therefore be like a zero-background experiment with only 0.02 photons per frame; the rate of useful frames in this comparison experiment, say having 2 or more photons, would only be 1 in 4000.

To test the reconstruction algorithm with data from randomly oriented samples, a pattern was cut out of x-ray opaque lead sheet to create an x-ray shadow mask (Figure~\ref{fig:realpadman1}). This mask was then mounted within a 19 mm diameter aperture of an opaque metal disk that fit onto a rotation stage (a Newport URS100BPP) with the center of rotation approximately at the center of the aperture. 

A very low-power copper anode x-ray tube was used to generate x-rays (TruFocus TFS 6050 Cu, 50 W maximum). It was operated at an anode voltage of 10.1 kV to reduce high-energy bremsstrahlung.  A 50 micron thick nickel filter was used to preferentially remove the $K_{\beta}$ of the tube spectrum to produce an approximately monochromatic x-ray beam of 8-keV Cu $K_{\alpha}$ radiation.  The rotation stage and aperture were mounted on the end of a 45 cm flight-path to produce a nearly flat-field illumination of x-rays across the 19 mm sample. 

The x-ray mask and a static x-ray image of the pattern are shown in Figures 1a and 1b. Cornell's LCLS Pixel Array Detector (PAD), comprising a single-chip 194 x 185 pixel array, was placed after the mask along the flight path. The PAD was positioned so the entirety of the aperture image was incident on the detector. The x-ray image shown in Figure 1b was collected by summing 432 images of the mask at fixed position. Each 0.1 s image had an occupancy of approximately 0.2 x-ray photons per pixel per frame in the unobstructed regions.

\begin{figure}[!t]
\includegraphics[width = \columnwidth]{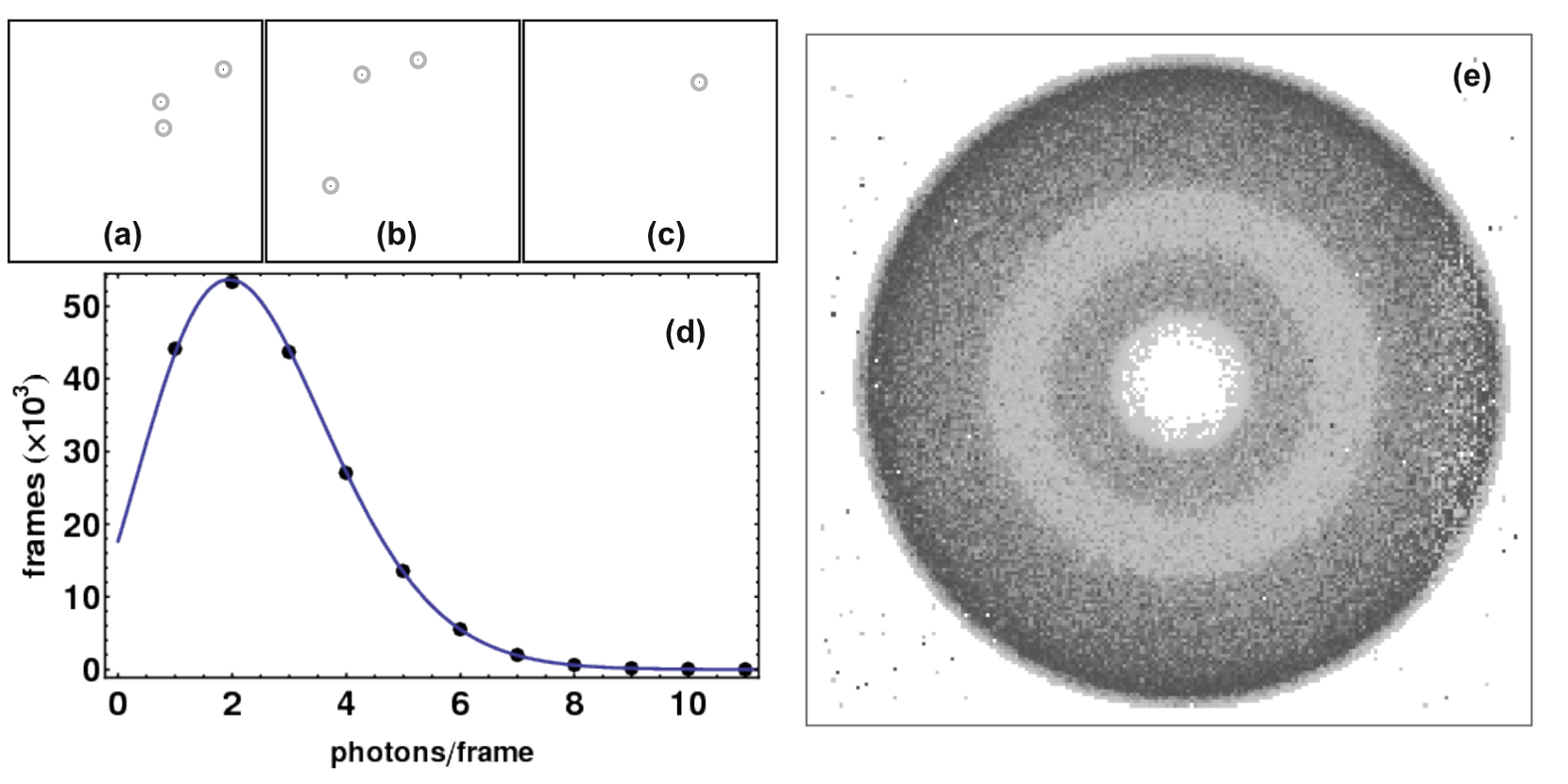}
\caption{(a-c) Three sample frames from the 2.5 photon/frame data set with detected x-ray photons circled. (d) Occupancy histogram compared with the Poisson distribution. (e) The sum of all thresholded frames from the 2.5 photon/frame data set showing a uniform angular distribution of data.}
\label{fig:lowand59dat}
\end{figure}
Very low-flux data were also collected with a continuously rotating sample. The detector collected images at 100 frames per second with a per-frame integration time of 100 microseconds. The waveforms used for digitization and detector readout were the same as those used when the detector is running at 120 frames per second, the frame rate of the LCLS, except the internal trigger was set to a 10 ms period (rather than 8 ms).  X-ray tube currents of 0.15 mA, and 0.03 mA were used to produce varying x-ray intensities.  With each current, hundreds of thousands of images were collected.  Figures 2(a-c) show three typical very low-flux mask images, in each case consisting of only a few x-rays per frame.

Dark signal measurements were made throughout the data collection sequence by periodically taking groups of 144 frames with the x-ray shutter closed. These dark frames were used to define a low-noise zero-level which was subtracted from individual signal frames to extract the x-ray induced signal from the raw detector output. 

Analog integrating detectors are required at XFELs because many other experiments deliver more than one x-ray photon per pixel per frame (as expected at low scattering angles in single particle imaging experiments), and the x-ray pulse is too short for photon counting electronics.  Minimum signal threshold values can be applied to the analog data to reject low-level noise~\cite{PhilippJinst1}. The threshold in this experiment was set to 0.7 x-ray photons (for 2.5 photon/frame data set) or 0.75 x-ray photons (for the 11.5 photon/frame data set).  At these thresholds approximately 0.05 and 0.01 false positive photon measurements per frame are expected, using a normal distribution and the previously measured~\cite{Philipp2010} pixel signal-to-noise ratio of 7 for a single 8-keV x-ray.  The lower threshold was used for 2.5 photon/frame data because this was the last data set taken and progressively less favorable parameters were chosen to test the robustness of the detector and algorithm.  No compensation was used for charge sharing between adjacent pixels, nor were pixel gains individually calibrated. A single, global threshold and nominal pixel gain value were applied across the array. 
 
Separate data sets analyzed included hundreds of thousands of frames with mean fluxes of 11.5 photons/frame and 2.5 photons/frame. 

Reconstructed images are shown in Figures~\ref{fig:115man} and~\ref{fig:25man}. Figure~\ref{fig:25man} was reconstructed using 450,000 frames of data with an average of 2.5 photons per frame. The reconstruction algorithm used 180 equally spaced $2^\circ$ steps. Figure 2e shows a simple sum of the thresholded frames of data that results in a rotationally smeared image with a uniform angular distribution.
This data set has a total of 1.2 million photons. For comparison, a data set with a similar total number of photons, but a higher per-frame photon average (and thus fewer frames) was also processed. The reconstruction is shown in Figure~\ref{fig:115man}, where the average occupancy was 11.5 photons/frame. 

The quality of the two reconstructions differs in both spatial resolution and contrast, with the 11.5 photons/frame data yielding better results. This agrees with the results of reconstructing 3D intensities from simulated single-particle diffraction data~\cite{loh026705}, also at very low flux. The degradation in quality occurs when a significant fraction of the information content in each frame, about half, is just the orientational state. There is a sharp increase in the iteration count of the EM algorithm when this criterion is met: the 2.5 photon/frame data required 220 iterations, compared to 49 iterations for the 11.5 photon/frame data.

\begin{figure}[h]
\subfloat[]{
\label{fig:25noback}
\includegraphics[width=0.45 \columnwidth]{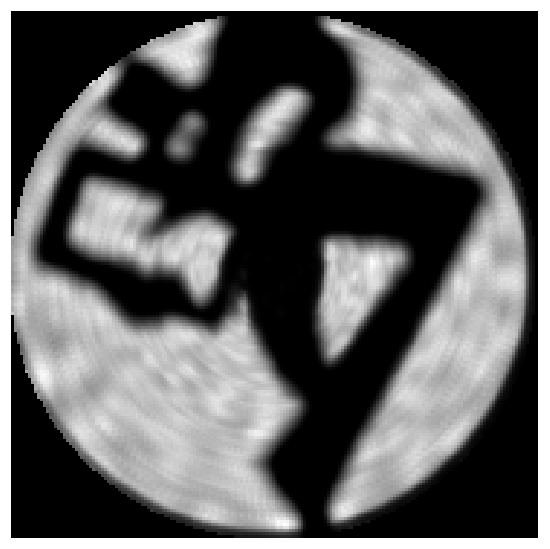}
}
\subfloat[]{
\includegraphics[width=0.45 \columnwidth]{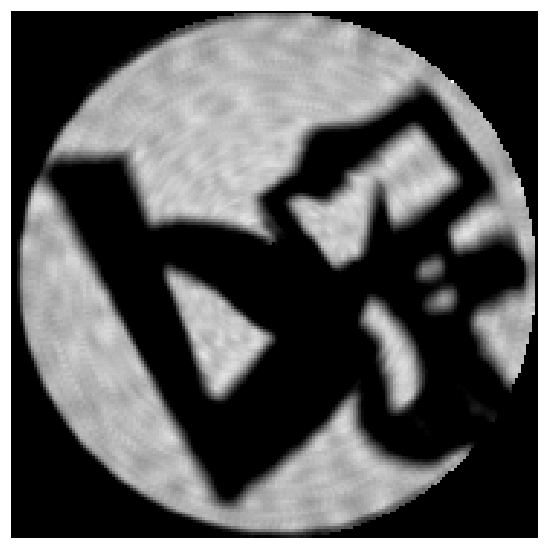}
}
\caption{\label{fig:115backsub} Effect of background on reconstruction quality.
(a) Reconstruction from 2.5 photons/frame data set and no added background. This is the same as Figure 1(d). 
(b) Reconstruction from the 11.5 photons/frame data set with an average of 11.5 photons of background added per frame `by hand' with a Poisson distribution. The background level was subtracted off before rendering to facilitate comparison to (a). As can be seen, the quality of the reconstructions is about the same, and much reduced from the original 11.5 photons/frame data (Figure 1(c)).
}
\end{figure}

By adding a uniform distribution of computer generated photon counts to the data sets, and processing it by the EM algorithm as before, we are able to simulate the effects of background scattering from gas molecules along the path of the incident x-ray beam in single particle experiments. This should be the major source of background signal and many times larger than the detector noise when the detector data are properly thresholded. Not surprisingly we find deterioration in the quality of the reconstruction. The degree of degradation is consistent with the information ratio $R$ quoted above, which equals 0.26 when signal-to-noise is 1. With this level of background our data set with 11.5 signal-photons/frame corresponds to a zero-background data set with only 3 photons/frame. The resulting reconstruction by the EM algorithm was therefore similar to that of our 2.5 photons/frame background-free reconstruction in both image quality and number of iterations (see Figure~\ref{fig:115backsub}).

Although this demonstration was motivated by the ongoing effort to realize single particle imaging, the strategy we employed applies more generally to measurements which seek to eliminate ensemble averaging and as a result yield extremely weak signals. Temporal averaging is avoided by short pulses of illumination and the spatial counterpart is achieved by isolation (e.g. single particles) or focusing, as in the case of ptychography~\cite{Thibault18072008}. In all these cases one sacrifices signal, thus putting an increased burden on the recording of weak signals with high fidelity and reconstructing from the resulting very sparse data. The envisioned single particle experiments at LCLS are an extreme example of this, but the same approach would apply even more to experiments with lower intensity sources, for example, Energy Recovery Linac (ERL) x-ray sources~\cite{Bilderback2010}.  An ERL can deliver very short x-ray pulses that are much less intense than XFEL pulses, but deliver many more pulses per second to compensate. Ptychography performed with an ERL, in conjunction with our data acquisition/analysis method, looks especially promising. Data acquisition would be fast and yet immune to mechanical instabilities because of the short pulse duration, while jitter in the position of the focus would be algorithmically reconstructed, in analogy with the angle reconstructions in our demonstration.

\section*{Acknowledgments}
LCLS PAD development was supported by subcontract from SLAC under DOE Contract DE-AC02-76SF00515.  Detector development at Cornell is also supported by DOE Grants FG02-97ER62443, DE-FG02-10ER46693 and the Keck Foundation.  CHESS is supported by NSF and NIH-NIGMS under NSF Grant DMR-0936384.  The data analysis dealing with the extraction from randomly-oriented, sparse data is supported by DOE Grant DE-FG02-11ER16210.

\end{document}